\def\year{2022}\relax
\documentclass[letterpaper]{article} 
\usepackage{aaai22}  
\usepackage{times}  
\usepackage{helvet}  
\usepackage{courier}  
\usepackage[hyphens]{url}  
\usepackage{graphicx} 
\usepackage{natbib}  
\usepackage{caption} 
\DeclareCaptionStyle{ruled}{labelfont=normalfont,labelsep=colon,strut=off} 
\usepackage{algorithm}
\usepackage{algorithmic}

\usepackage{newfloat}
\usepackage{listings}
\floatstyle{ruled}
\newfloat{listing}{tb}{lst}{}
\floatname{listing}{Listing}
\usepackage{amsmath,amssymb,amsfonts}

\usepackage{array}
\usepackage{booktabs}
\usepackage{graphicx}
\usepackage{threeparttable}
\usepackage{wasysym}

\usepackage[framemethod=tikz]{mdframed}

\makeatletter
\def\ProblemSpecBox{
  \@ifnextchar[\ProblemSpecBox@opt{\ProblemSpecBox@noopt}}
\def\ProblemSpecBox@opt[#1]#2{
  \protected@edef\@currentlabelname{#1}
  \protected@edef\@currentlabel{#1}
  \begin{mdframed}[
    innerlinewidth=0.5pt,
    innerleftmargin=10pt,
    innerrightmargin=10pt,
    innertopmargin = 10pt,
    innerbottommargin=10pt,
    skipabove=\dimexpr\topsep+\ht\strutbox\relax,
    roundcorner=5pt,
    frametitle={#2},
    frametitlerule=true,
    frametitlerulewidth=1pt]
}
\def\ProblemSpecBox@noopt#1{
  \ProblemSpecBox@opt[#1]{#1}
}
\def\endProblemSpecBox{
  \end{mdframed}
}
\makeatother

\title{\textsc{\mdseries Scotch}: An Efficient Secure Computation Framework for Secure Aggregation}

\author {
    Yash More\textsuperscript{\rm 1},
    Prashanthi Ramachandran\textsuperscript{\rm 2},
    Priyam Panda\textsuperscript{\rm 1},
    Arup Mondal \textsuperscript{\rm 1},
    Harpreet Virk\textsuperscript{\rm 1},
    Debayan Gupta\textsuperscript{\rm 1}
}
\affiliations {
        \textsuperscript{\rm 1} Ashoka University \\
        \textsuperscript{\rm 2} Brown University \\
         yash.more@alumni.ashoka.edu.in, pramach3@cs.brown.edu, priyam.panda\_asp22@ashoka.edu.in, 
         arup.mondal\_phd19@ashoka.edu.in,
         harpreet.virk@alumni.ashoka.edu.in, debayan.gupta@ashoka.edu.in
}

\usepackage{bibentry}

\begin{document}

\maketitle

\begin{abstract}
Federated learning enables multiple data owners to jointly train a machine learning model without revealing their private datasets. 
However, a malicious aggregation server might use the model parameters to derive sensitive information about the training dataset used. To address such leakage, differential privacy and cryptographic techniques have been investigated in prior work, but these often result in large communication overheads or impact model performance. To mitigate this centralization of power, we propose \textsc{Scotch}, a decentralized \textit{m-party} secure-computation framework for federated aggregation that deploys MPC primitives, such as \textit{secret sharing}. Our protocol is simple, efficient, and provides strict privacy guarantees against curious aggregators or colluding data-owners with minimal communication overheads compared to other existing \textit{state-of-the-art} privacy-preserving federated learning frameworks. We evaluate our framework by performing extensive experiments on multiple datasets with promising results. \textsc{Scotch} can train the standard MLP NN with the training dataset split amongst 3 participating users and 3 aggregating servers with 96.57\% accuracy on MNIST, and 98.40\% accuracy on the Extended MNIST (digits) dataset, while providing various optimizations. 
\end{abstract}

%
%
\section{Introduction}\label{sec:intro}

Standard machine learning environments often rely on large amounts of sensitive data to achieve a high level of performance~\cite{35179}. However, preparing a central repository of data is laborious, making secure-collaborative training expensive. Outsourcing the data to a central server that performs model training for the users is a potential solution, but is often not feasible in privacy-sensitive settings. 
Secure aggregation of data using multiparty computation frameworks (MPC)~\cite{yao1982protocols,mood2016frigate,perry2014systematizing,di2014practical,gupta2016using,lindell2020secure,goldreich1998secure,goldreich2019play} has been explored in recent works, but they significantly impact framework efficiency due to added computational overheads~\cite{phong}. Moreover, centralized aggregation creates a single point of failure in the framework that can potentially compromise the security and privacy of the training data if the server is malicious, or prone to adversarial attacks by colluding participants ~\cite{chen2021robust,kairouz2021advances}. 

A recently proposed alternative for privacy-preserving training, without data outsourcing, is \textit{Federated Learning} (FL)~\cite{mcmahan2016communication}. FL has emerged as a promising approach to collaboratively train a model by exchanging model parameters with a central aggregator (or server), instead of the actual training data. However, parameter exchange may still leak a significant amount of private data~\cite{zhu}. Several approaches have been proposed to overcome this leakage problem based on differential privacy (DP)~\cite{shokri2015privacy,papernot2018scalable}, MPC~\cite{bonawitz2017practical,ryffel2018generic}, HE~\cite{truex2019hybrid}, etc. While DP-based learning aims to mitigate inference attacks, it significantly degrades model utility as the training of accurate models requires high privacy budgets ~\cite{jayaraman2019evaluating}.
Cryptographic techniques provide improved privacy protection but remain too slow for practical use due to the extensive cryptographic operations. Hence, there arises a need for a secure, decentralized FL framework that protects user privacy, while allowing seamless training of ML models. This requires strong cryptographic protection of the intermediate model updates during the model aggregation and the final model weights. 

In this work, we propose \textsc{Scotch}, a \textit{practical} framework that enables secure \textit{m-party} aggregation in a distributed \textit{n-server} setting. It provides end-to-end protection of the parties’ training data, intermediate model weights, and the final resulting model by combining secure multiparty computation (MPC) primitives based on \textit{secure outsourced computation} and \textit{secret sharing} to enable decentralized FL. Our contributions have been described in further detail in the following section.

\subsection{Our Contributions}

In this paper, we introduce a one-of-its-kind framework for privacy-preserving federated learning with primitives from conventional machine learning and multiparty computation (MPC). Specifically, 

\begin{itemize}
    \item  We propose \textsc{Scotch}, a simple, fast, and efficient federated learning framework that allows for decentralized gradient aggregation using \textit{secure outsourced computation} and \textit{secret sharing} while ensuring strict privacy guarantees of the training data ~\cite{mohassel2017secureml,wagh2019securenn}.
    
    \item We evaluate the efficiency of our proposed \textit{secret sharing}-based FL protocol against existing \textit{state-of-the-art} frameworks. To the best of our knowledge, \textsc{Scotch} is the only approach for decentralized privacy-preserving FL with the least possible cryptographic computational overheads -- only $O(2mn)$ crypto-related operations required in each training round, where $m$ is the number of participants and $n$ is the number of aggregators (See Table \ref{table:complexity}).
    
    \item We implement \textsc{Scotch} and perform extensive experiments on multiple standard datasets such as MNIST, EMNIST, and FMNIST with promising results: \textsc{Scotch} has efficiency improvements both in training time and communication cost while providing similar model performance and privacy guarantee as other approaches. 
\end{itemize}

    


For ease of access, all of our code and experiments are available at: \textcolor{blue}{\url{https://github.com/arupmondal-cs/SCOTCH}}.

\section{Technical Background}\label{sec:techback}

\paragraph{Federated Learning.}
FL~\cite{mcmahan2016communication} is a distributed ML approach that enables model training on a large corpus of decentralized data with myriad participants. It is an example of the more general approach of ``bring code to data, not data to code''. In FL, each party trains a model locally and exchanges only model parameters with an FL \textit{server} or \textit{aggregator}, instead of the private training data.

The participants in the training processes are \textit{parties} and the \textit{FL server}, which is a cloud-based distributed service. Devices agreement to the server that they are ready to run an \textit{FL task} for a given \textit{FL population}. An FL population is specified by a globally unique name which identifies the learning problem, or application, which is worked upon. An FL task is a specific computation for an FL population, such as training to be performed with given hyperparameters, or evaluation of trained models on local device data. After finishing the local computation on its local dataset then each device updates the model parameters (e.g. the weights of a neural network) to the FL server. The server incorporates these updates into its global state of the global model.

\paragraph{Secure Multiparty Computation.}

Secure multiparty computation (MPC)~\cite{yao1982protocols,mood2016frigate,perry2014systematizing,di2014practical,gupta2016using,lindell2020secure,goldreich1998secure,goldreich2019play} is the universal cryptographic functionality, allowing any function to be computed obliviously by a group of mutually distrustful parties. There exist a number of different techniques for MPC (\textit{e.g.,} garbled circuits, functional encryption, and homomorphic encryption, etc.). In this work, we have considered MPC based on secret sharing~\cite{shamir1979share}.

\paragraph{Secret Sharing.}
In cryptography, secret sharing~\cite{shamir1979share,blakley1979safeguarding} refers to the process of splitting a secret among $n$ parties such that each party does not learn anything about the whole secret from the share it holds. The secret can be reconstructed only if a certain minimum number of parties, greater than or equal to a threshold, $t$, combine their shares. The scheme is known as the $(t,n)$ threshold scheme or $t$-out-$n$ secret sharing. In this work, we use additive secret sharing, which uses addition as the way to combine shares. We use the notation $[a]_j$ to denote the $j^{th}$ share of a secret $a$. 


\section{Proposed Framework}\label{sec:frame}

\begin{figure}[ht]
    \centering
    \includegraphics[width=0.38\textwidth]{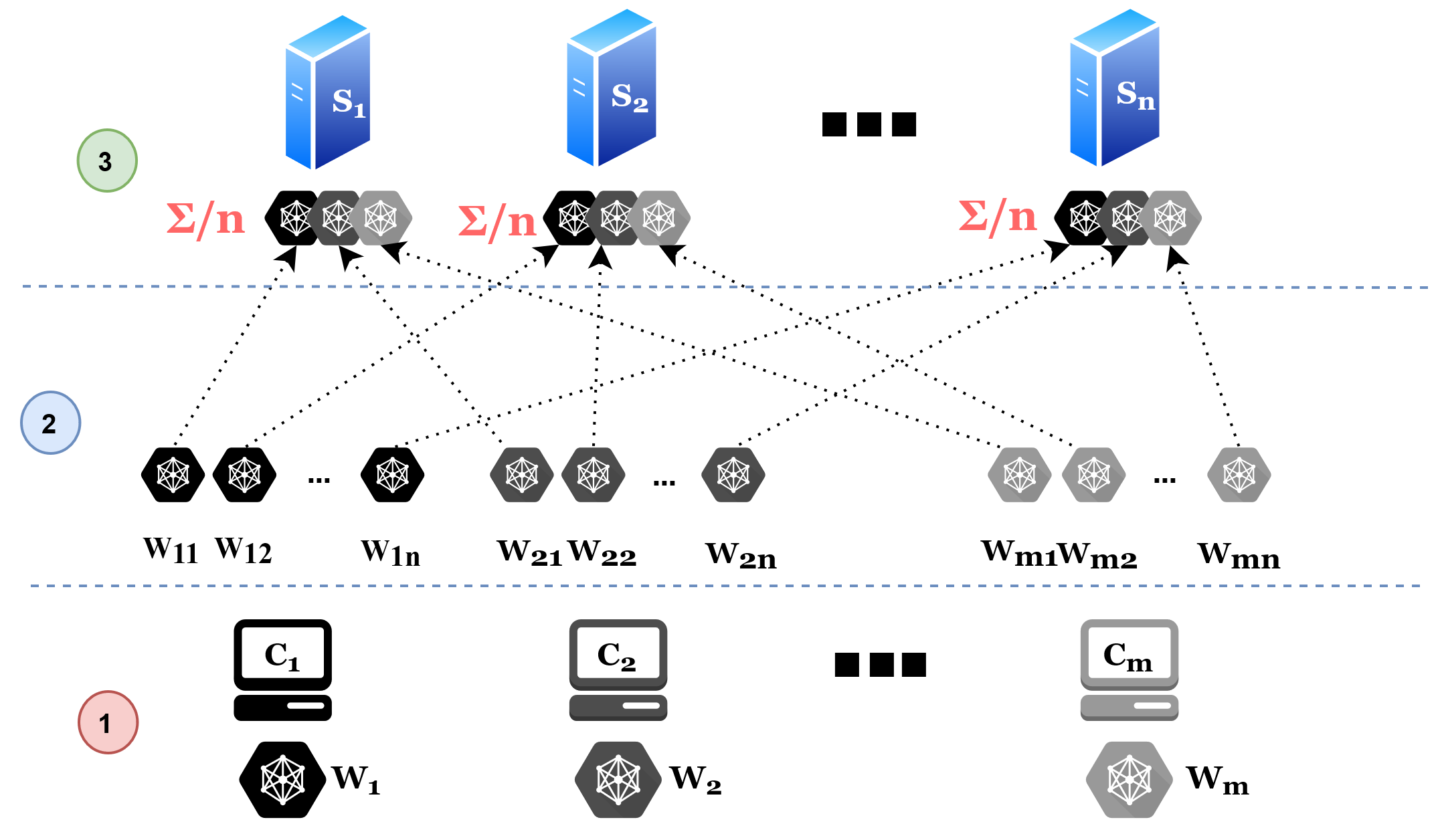}
    \caption{Schematic Diagram illustrating Algorithm~\ref{alorithm:flframe}.}
    \label{fig:ppfldig}
\end{figure}

In this section, we describe the proposed framework \textsc{Scotch}, an efficient distributed secure-computation approach for secure outsourced aggregation based on MPC primitives. The distributed federated averaging algorithm has been described in Algorithm~\ref{alorithm:SCOTCHframe}. Algorithm~\ref{alorithm:flframe} briefly describes one iteration of our protocol. The steps given in this algorithm have been illustrated in Figure~\ref{fig:ppfldig}. 

\subsection{Threat Model}\label{subsec:thmod}
We assume a passively secure threat model. A \textit{passive} (honest-but-curious) adversary follows the protocol specifications but may try to learn information about the private input data by inspecting the shared inputs. Both the participants, data owners (or clients), and the aggregators (or servers) are \textit{honest-but-curious}. \textsc{Scotch} ensures that aggregators (collude with any subset of participants and aggregators) can't learn any information about the private inputs of the honest participants. Similarly, it also ensures that any subset of colluding participants cannot learn any information about the private inputs or outputs of the honest participants by inspecting the messages exchanged with the aggregators or the final model. We also assume any encryption broadcast to the network in Algorithm~\ref{alorithm:flframe} is re-randomized to avoid leakage about parties’ confidential data by two consecutive broadcasts. We omit this operation in Algorithm~\ref{alorithm:flframe} for clarity. Finally, \textit{attacks that aim to create denial-of-service attacks or inject malicious model updates are beyond the scope of this short paper}.

\subsection{\textsc{\mdseries Scotch} Framework}

We assume a set of $n$ honest-but-curious aggregators, $\mathcal{S}$ and a set of $m$ clients, $\mathcal{C}$, where each client $\mathcal{C}_i$ for $i \in \{1,\cdots,m\}$ holds its own private dataset $\mathcal{D}_i$. We defer more details about the threat model and security of the framework to `threat model' and `privacy guarantees' section. The clients in $\mathcal{C}$ agree upon a model architecture, $\text{NN}_{\text{arch}}$, for local training prior to the runtime of the framework. The underlying concept in this framework is $n$-out-of-$n$-additive-secret-sharing-based MPC, which provides protocols for $n$ aggregators and is secure against a passive adversary that corrupts at most $m-1$ clients.

\paragraph{Local Training.}
At the beginning of every iteration, the function \texttt{local\_training} is invoked by client $\mathcal{C}_i$ in $\mathcal{C}$ with input $\mathcal{D}_i$. This function allows clients to train local models on their private datasets using the pre-decided model architecture, $\text{NN}_\text{arch}$. In the first iteration, initial weights are sampled and stored in $w$. For subsequent iterations, the aggregated weights from the previous iteration are used as initial weights. Each client samples a randomly-permuted (without replacement) subset $d$ from the dataset $\mathcal{D}_i$ in each iteration. Functions \texttt{permute\_indices} and \texttt{choose\_subset} help with the same. In each iteration, each client trains a model on $\text{NN}_\text{arch}$ with inputs $w$ and $d$. The clients then split the model weights into $n$-out-of-$n$ additive secret shares by invoking \texttt{split\_secret\_shares}. These shares are then sent to the $n$ aggregators. 

\paragraph{Secure Aggregation.}
Having received a total of $m$ shares from clients in $\mathcal{C}$, each server $\mathcal{S}_j$ for $\{1,\cdots,n\}$ adds its local shares and divides the sum by the total number of aggregators to obtain the value $\sigma_j$ by invoking \texttt{federated\_sum}. One can observe that $\sigma_j$ is an $n$-out-of-$n$ additive secret share of the federated average of the local models of the clients. Each server then sends $\sigma_j$ to clients in $\mathcal{C}$ so that they can obtain the final model. 

\paragraph{Computing the Final Model.}
Having obtained additive secret shares of the federated average from the $n$ aggregators, each client locally adds up the shares to obtain the federated average of their models by invoking the function \texttt{compute\_final\_model}. Clients set the value of variable $w$ as the federated average obtained in this iteration. If the current iteration is the final one, $w$ is returned as the final output. If not, $w$ is used as the initial weights in \texttt{local\_training} for the subsequent iteration.

To enable seamless integration between machine learning primitives (which generally use floating-point), and MPC primitives (which generally use integers), we use integer ring arithmetic in our implementation. To enable conversions between the float and integer realms, we use functions \texttt{float\_to\_int}, \texttt{int\_to\_float}, and \texttt{truncate} based on primitives provided in~\cite{mohassel2017secureml}. After training its local model, each client embeds its weights onto the integer ring by invoking \texttt{float\_to\_int}\footnote{\texttt{float\_to\_int} converts a floating-point value into an $l$-bit integer by allocating $l_x$ bits to the integer part, $l_f$ bits to the fractional part, and $1$ bit to the sign of the value, such that $ l = l_x + l_f + 1$. Note that $l_f$ represents the maximum \textit{precision} of the value. Refer to~\cite{wagh2019securenn} for further details.}. The rest of the operations are performed in the integer ring realm. At places where two values in the integer ring are multiplied, the product is truncated by invoking \texttt{truncate}. Finally, at the end of every iteration, the aggregated weights are converted back to float by invoking \texttt{int\_to\_float} in order to facilitate any further local training on them.


  
  
  
  
  
  

\begin{ProblemSpecBox}[1]{Algorithm 1: Secure Outsourced Aggregation\label{alorithm:flframe}}
    \textbf{Input:}  Client $\mathcal{C}_i$ for $i \in \{1,\cdots,m\}$ holds its private dataset $\mathcal{D}_i$. \newline
    \textbf{Output:} Client $\mathcal{C}_i$ for $i \in  \{1,\cdots,m\}$ obtains the final aggregated global model, $\mathbb{M}_{\text{agg}}$. 
    
    \begin{enumerate}
        \item Client $\mathcal{C}_i$ for $i \in \{1,\cdots,m\}$ trains local model $\mathbb{M}_i$ on a random subset of its private dataset $\mathcal{D}_i$. Note that all the clients use the same model architecture.
  
        \item Client $\mathcal{C}_i$ for $i \in \{1,\cdots,m\}$ creates $n$ additive secret shares of its model $\mathbb{M}_{i}$ and sends each share $\mathcal[{M}_{i}]_{j}$ for $j \in  \{1,\cdots,n\}$ to server $\mathcal{S}_j$. 
  
        \item Server $\mathcal{S}_j$ for $j \in  \{1,\cdots,n\}$ adds up the received shares from all clients and divides the sum by $n$ to obtain $\sigma_j$ and sends $\sigma_j$ to all the clients. \newline
        Each client locally computes $\mathbb{M}_{\text{agg}} = \sum_{j=1}^{n} (\sigma_j$). 
    \end{enumerate}

\end{ProblemSpecBox}

\begin{ProblemSpecBox}[2]{Algorithm 2: \textsc{ Scotch} Framework\label{alorithm:SCOTCHframe}}

    \textbf{Input:} Client $C_i$ in $\mathcal{C}$ possesses private dataset $\mathcal{D}_i$ for $i \in \{1,\cdots, m\}$. $iter :=$ the total number of global iterations for aggregation len($\mathcal{D}_i)$ represents the number of data points in the dataset $\mathcal{D}_i$. $n$ is the total number of aggregators.\\
    \textbf{Output:} Clients obtain the final aggregated model stored in $w$.
    
    \begin{itemize}
        \item[-] \textbf{foreach} {$k \in \{1, \cdots, iter \}$}
        \begin{itemize}
            \item[-] \textbf{foreach}  {$i \in \{1,\cdots,m\}$}
            \begin{itemize}
                \item[-] \texttt{local\_training}($\mathcal{C}_i$, $\mathcal{D}_i$); \\ 
            \end{itemize}

            \item[-] \textbf{Procedure} \texttt{local\_training} ($\mathcal{C}_i$, $\mathcal{D}_i$):
            \begin{itemize}
                \item[-] \textbf{If}($k == 1$)
                \begin{itemize}
                    \item[-] $w \leftarrow$ \texttt{random\_init()}; {\scriptsize // randomly sample initial weights for n.}
                \end{itemize}
                \item[-] $d = $\texttt{permute\_indices}($\mathcal{D}_i)$;
                \item[-] $d = $\texttt{choose\_subset}($d$, len($D_i$) / $iter$);
                \item[-] $W_i \leftarrow $ train($w$, $d$);
                \item[-] $\mathbb{M}_i \leftarrow $ \texttt{float\_to\_int}($W_i$);
                \item[-] $ \{[\mathbb{M}_i]_1, \cdots, [\mathbb{M}_i]_n\} \leftarrow \texttt{split\_secret\_shares}(\mathbb{M}_i)$; {\scriptsize // split the model into $n$-out-of-$n$ additive secret shares.} \\
            \end{itemize}

            \item[-] \textbf{foreach} {$j \in \{1,\cdots,n\}$}
            \begin{itemize}
                \item[-] \texttt{federated\_sum}($S_j$, $[\mathbb{M}_i]_j$); \\
            \end{itemize}

            \item[-] \textbf{Procedure} ($S_j,\{ [\mathbb{M}_1]_j, \cdots, [\mathbb{M}_m]_j\}$):
            \begin{itemize}
                \item[-] $\sigma_j \leftarrow \sum_{i=1}^{m} ([\mathbb{M}_i]_j) \times$ \texttt{float\_to\_int}($1/n$);
                \item[-] $\sigma_j \leftarrow$ \texttt{truncate}($\sigma_j$); \\
            \end{itemize}
            
            \item[-] \textbf{foreach} {$i \in \{1,\cdots,m\}$}
            \begin{itemize}
                \item[-] \texttt{compute\_final\_model}($\mathcal{C}_i$, $\sigma_j$); \\
            \end{itemize}
            
            \item[-] \textbf{Procedure} ($\sigma_1,\cdots,\sigma_n$):
            \begin{itemize}
                \item[-] $\mathbb{M}_{agg} \leftarrow \sum_{j=1}^{n} (\sigma_j)$;
                \item[-] $w \leftarrow$ \texttt{int\_to\_float}$(\mathbb{M}_{agg})$; \\
            \end{itemize}
        \end{itemize}
        \item[-] \textbf{return} $w$;
    \end{itemize}
\end{ProblemSpecBox}

\subsubsection{Communication Complexity}

Table~\ref{table:complexity} describes the complexity of the secure aggregation protocol (refer to Algorithm~\ref{alorithm:SCOTCHframe}). Since \textsc{Scotch} is a secure aggregation framework, the complexity of functions \texttt{local\_training} and \texttt{compute\_final\_model} can be considered \textit{offline}. As a result, we only consider \texttt{federated\_sum()} as the \textit{online} phase of the protocol. 

\begin{table}[ht]
\centering
\caption{Summary of the complexity of Algorithm~\ref{alorithm:SCOTCHframe}.}
\begin{tabular}{|l|c|c|}
\hline
\multicolumn{1}{|l|}{Complexity} & Data Owners & Aggregator Servers \\ \hline
Computation            & $O(2mn)$ & $O(mn)$ \\ \hline
Communication          & $O(n)$  & $O(m)$  \\ \hline
Storage                & $O(m)$  & $O(n)$  \\ \hline
\end{tabular}
\label{table:complexity}
\end{table}

\subsection{Privacy Guarantees}\label{sec:secpriv}
\textsc{Scotch} achieves data privacy guarantees under a semi-honest adversary model with any subset of colluding aggregators. \textsc{Scotch}'s infrastructure is designed using multi-input \textit{secret sharing}-based MPC protocol to calculate the federated average of model gradients shared by participating clients. Private training data is not sent -- participating entities get split ``shares'' of model gradients, or the generated averaged model, neither of which can be used to reconstruct sensitive information about the training dataset used. The security of these shares is guaranteed by standard MPC theorems~\cite{goldreich1998secure}, and since the actual computations performed within the MPC setup (which can perform arbitrary computations and is agnostic in that sense). 

\section{Experimental Evaluation}\label{sec:exp}

\subsection{Implementation Details}
We simulate \textsc{Scotch} using socket, a low-level networking interface that can be accessed using Python. We rely on the Tensorflow library for the training and inference of machine learning models. All our experiments are performed on a local machine -- a Linux machine with Intel i7-9700K CPU@3.60 GHz and GeForce RTX 2070 GPU with 32 GB RAM. All clients and servers are assumed to be running independent nodes and are connected via a virtual network.

\subsection{Dataset and Model Configuration}\label{subsec:datamodelconfig}

\subsubsection*{MNIST~\cite{mnist}.}
\begin{figure}[ht]
\centering
\includegraphics[width=0.4\textwidth]{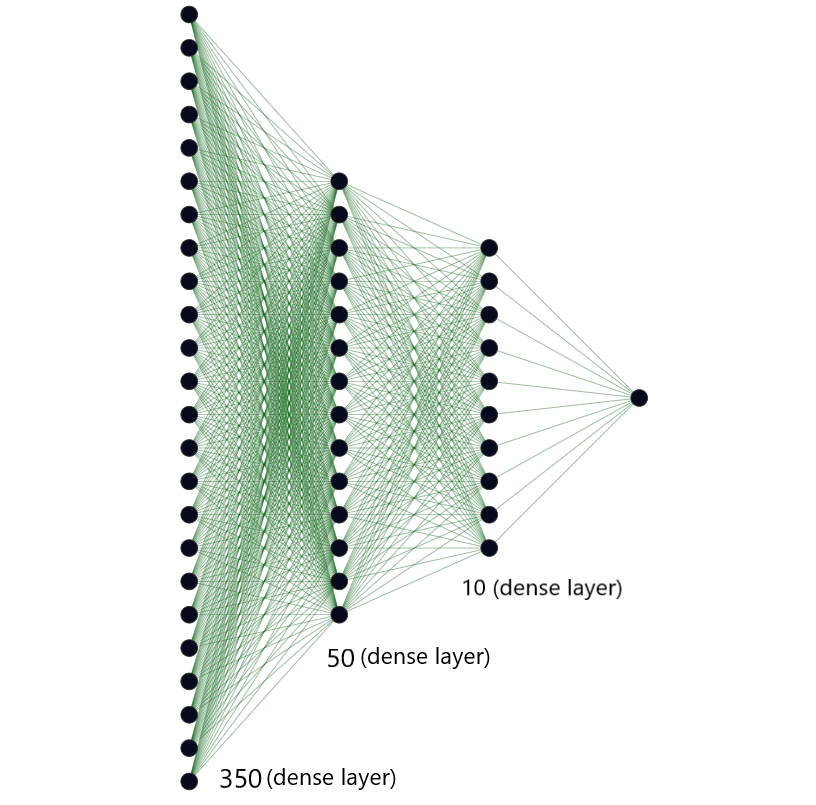}
\caption{Multilayer Perceptron (MLP) architecture used across different experiments in \textsc{Scotch}.}
\label{fig:mlp_arch}
\end{figure}

The MNIST~\cite{mnist} dataset comprises 60,000 handwritten digit character images, along with 10,000 testing images. The data is pre-processed by resizing each image, and one-hot encoding the labels. Each client uses a three-layer Multi-layered Perceptron to train on their local datasets. The architecture of the MLP is outlined in Figure~\ref{fig:mlp_arch}.

\subsubsection*{EMNIST~\cite{emnist}.}
The Extended MNIST (digits)~\cite{emnist} dataset contains 240,000 handwritten digit character images and 40,000 images for training and testing purposes respectively. The data is preprocessed by resizing each image, and one-hot encoding the labels. We use the same MLP architecture as used in MNIST, to train each local model.

\subsubsection*{FMNIST~\cite{fmnist}.}
Fashion-MNIST~\cite{fmnist} is a dataset of Zalando's article images that contains 60,000 training images and 10,000 testing images. The data is pre-processed by resizing each image, and one-hot encoding the labels.


\subsection{Experimental Overview}
\textsc{Scotch}'s framework incorporates secure aggregation via \textit{secret outsourced computation}. Each client takes part in federated learning by (a) locally training on their private data, and (b) sharing their gradients with servers via secret-sharing. Each server receives partial shares from the clients, which it aggregates and propagates back to all clients. This allows each client to recompute the global model gradients by averaging the shares received from the server(s).

\subsection{Experimental Results}

We evaluate \textsc{Scotch} in terms of three indicators: (a) Performance of the generated model with a different number of clients and servers, (b) Impact of varying precision while \textit{secret sharing} on model performance, and (c) communication complexity (see Table~\ref{table:complexity}).

\subsubsection{Performance Analysis.}
We evaluate \textsc{Scotch}'s performance on three standard datasets -- MNIST, EMNIST, and FMNIST (refer Dataset and Model Configuration section) with varying numbers of clients, in a 3-server setting. For each dataset, we use a three-layer MLP whose architecture has been outlined in \S\ref{subsec:datamodelconfig}. We use a standard 70-30 train-test split, for each dataset, and the training data is equally divided amongst the clients. Each client locally trains on their individual dataset for 3-4 epochs, with a learning rate of 0.01. The results have been summarized in Table~\ref{table:SCOTCH_acc}. 

To test the effects of precision on training our global model, we compare the results of \textsc{Scotch} on MNIST dataset, with 16 and 32 bits of precision. The test accuracy comparison between these two is shown in Table~\ref{table:Precision}. To support decimal arithmetic in an integer ring we use the solution proposed by~\cite{mohassel2017secureml}. As we observe from our experiments, if we restrict the number of decimal places to 32 bits, we see a significant improvement in test accuracy as opposed to 16 bits. Therefore, we observe a direct correlation between the precision of floating-point numbers involved in network training and the resulting model. 
To understand the effects of precision, we trained a centralized FL server with a constraint -- we round each weight update of the ML model with 32 bits of precision (restricting values up to 5 decimal places). We observed that there is a considerable decrease in model accuracy with decreasing precision. This underscores the importance of precision while training machine learning models. We summarize our observations in Table~\ref{table:Centralized} (for further details, please refer to the Impact of Precision Length section). 

We observe a \textit{decrease in accuracy with increasing number of clients} because of the compounding errors in \texttt{float\_to\_int()} \textit{and} \texttt{int\_to\_float()} conversions as a result of \textit{limited precision}. These can be offset by an increase in precision.
We plan to scale our existing framework to a larger number of clients and servers with the help of a reasonable increase in precision size in the near future.

\begin{table}[ht]
    \centering
    \caption{\textsc{Scotch}'s performance accuracy on MNIST~\cite{mnist}, under 16-bit and 32-bit precision configuration. The number of global iterations for aggregation,  \textit{iter} (Algorithm~\ref{alorithm:SCOTCHframe}) is set to $4$. For the accuracy graph, see Figure~\ref{fig:SCOTCH_plot_acc_pre}.}
    \label{table:Precision}
    \begin{tabular}{|l|l|l|l|l|l|}
    \hline
    Clients & 2 & 3 & 4 & 5 \\ \hline
    MNIST-16 & 0.3 & 0.19 & 0.113 & 0.111  \\ \hline
    MNIST-32 & 0.975 & 0.965 & 0.74 & 0.53  \\ \hline
    \end{tabular}
\end{table}

\begin{table}[ht]
    \caption{\textsc{Scotch} performance accuracy, as evaluated on three datasets: MNIST, EMNIST, and FMNIST with an increasing number of clients (under 32-bit precision). For the accuracy graph, see Figure~\ref{fig:SCOTCH_plot_acc}.}
    \label{table:SCOTCH_acc}
    \centering
    \begin{tabular}{|l|l|l|l|}
    \hline
        Clients & MNIST & EMNIST& FMNIST \\ \hline
        2 & 0.975 & 0.985 & 0.85 \\ \hline
        3 & 0.965 & 0.984 & 0.69 \\ \hline
        4 & 0.74 & 0.9 & 0.53 \\ \hline
        5 & 0.53 & 0.549 & 0.5 \\ \hline
    \end{tabular}
\end{table}

\begin{table}[ht]
    \caption{Evaluating the performance accuracy of Centralized FL (1-server setting) on multiple precision configurations using the MNIST~\cite{mnist} dataset.}
    \label{table:Centralized}
    \centering
    \begin{tabular}{|l|l|l|l|l|}
    \hline
        Precision & 4-bit & 8-bit & 16-bit & 32-bit \\ \hline
        Centralized FL & 0.09 & 0.41 & 0.71 & 0.85 \\ \hline
    \end{tabular}
\end{table}

\begin{figure}[ht]
\centering
\includegraphics[width=0.45\textwidth]{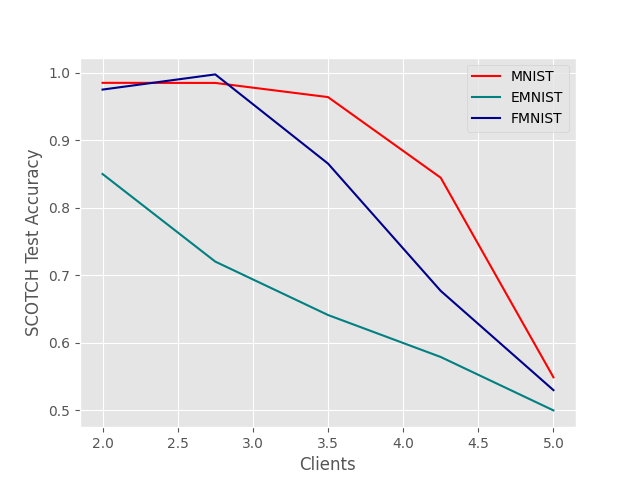}
\caption{\textsc{Scotch}: Clients vs Performance accuracy on multiple datasets -- MNIST, EMNIST, and FMNIST (3-server setting).}
\label{fig:SCOTCH_plot_acc}
\end{figure}

\begin{figure}[!ht]
\centering
\includegraphics[width=0.45\textwidth]{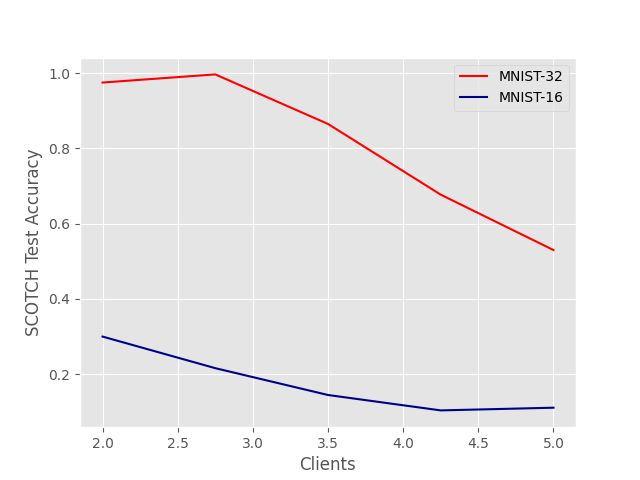}
\caption{\textsc{Scotch}: Clients vs Performance accuracy on multiple precision settings (16-bits, 32-bits) on MNIST (3-server setting).}
\label{fig:SCOTCH_plot_acc_pre}
\end{figure}

\subsection{Impact of Precision Length}

Most protocols in secure multiparty computation operate in integer rings. However, one needs to deal with decimal numbers while tackling computations in machine learning algorithms. 
To mitigate this, we use a mapping between fixed-point decimals and the integer ring (as used by state-of-the-art MPC frameworks such as SecureML~\cite{mohassel2017secureml}). The integer part of the decimal number is represented by $l_w$ bits and the fractional part by $l_f$. 
To evaluate the effects of precision on our training and testing accuracy, we replicate the precision settings used in SecureML~\cite{mohassel2017secureml} for logistic and linear regression. 
Comparing our tests with SecureML helps us to understand the effects of precision-length while training different machine learning models.
Even though SecureML's experiments were restricted to 13--16 bits, they used  much simpler models such as logistic regression and  with a simpler dataset -- (1000 to 1M samples of the MNIST dataset), and objective -- Binary Classification. 
Through our experiments we observe that multi-class classification via Multi-layer Perceptrons on much smaller dataset 70\% of 60,000 MNIST images performs better if we increase the precision (refer to Table~\ref{table:Centralized}). One might notice how much of a difference does precision make on gradient updates while performing gradient descent using Neural Networks. This difference plays a role in our experiments as well.
\section{Related Work}\label{sec:rlwork}

The existing privacy-preserving machine learning (PPML) works focus exclusively on training (generalized) linear models. They rely on \textit{centralized} solutions where the learning task is securely outsourced to a server, notably using homomorphic encryption (HE) techniques. As such, these works do not solve the problem of privacy-preserving distributed ML, where multiple parties collaboratively train an ML model on their data. To address the latter, several works propose multi-party computation (MPC)~\cite{yao1982protocols,mood2016frigate,perry2014systematizing,di2014practical,gupta2016using,lindell2020secure,goldreich1998secure,goldreich2019play} solutions where several tasks, such as clustering and regression, are distributed among 2, 3, or 4 servers~\cite{mohassel2017secureml,wagh2019securenn,blaze,ramachandran2021s++,wagh2021falcon,riazi2018chameleon,demmler2015aby,payman2018aby3,wagh2021falcon}. Although such approaches, however, limit the number of parties among which the trust is split, often assume an honest majority among the computing servers, and require parties to communicate (i.e., secret share) their data outside their premises. This might not be acceptable due to the privacy and confidentiality requirements and the strict data protection regulations. 

A recently proposed alternative for privacy-preserving training -- without data outsourcing -- is \textit{federated learning} (FL)~\cite{mcmahan2016communication}. FL has emerged as a promising approach to collaboratively train a model by exchanging model parameters with a central aggregator, instead of the actual training data. However, parameter exchange may still leak a significant amount of private data. Several approaches have been proposed to overcome this leakage problem based on differential privacy (DP)~\cite{shokri2015privacy,papernot2018scalable}, MPC~\cite{bonawitz2017practical,ryffel2018generic}, HE~\cite{truex2019hybrid,sav2020poseidon}, Trusted Execution Environment~\cite{mondal2021flatee,mondal2021poster}, etc. Furthermore, in those settings, the aggregator is a central player, which also potentially represents a single point of failure~\cite{kairouz2021advances} and due to the extensive use of cryptographic operations, these frameworks remain too slow for practical use. Finally, other works combine MPC with DP techniques to achieve better privacy guarantees~\cite{truex2019hybrid,xu2019hybridalpha,pettai2015combining}. While DP-based learning aims to mitigate inference attacks, it significantly degrades model utility, as training accurate models requires high privacy budgets~\cite{jayaraman2019evaluating}. Therefore, a \textit{practical} distributed privacy-preserving federated learning approach requires strong cryptographic protection of the intermediate model updates during the model aggregation and the final model weights. 
\section{Conclusion}\label{sec:conclusion}

We propose \textsc{Scotch}, a decentralized \textit{m-party}, \textit{n-server} secure-computation framework for federated aggregation that utilizes MPC primitives. The protocol provides strict privacy guarantees against honest-but-curious aggregators or colluding data-owners; it offers the least communication overheads compared to other existing \textit{state-of-the-art} privacy-preserving federated learning frameworks on standard datasets. In the near future, we plan to extend this framework to provide security against malicious servers and clients, scale it to a larger number of clients and servers, and finally deploy it via open-source channels for academic and industrial use-cases.

%
%
\bibliography{ms}

%
%

\end{document}